\documentclass[aps,pre,onecolumn,superscriptaddress,longbibliography,nobibnotes,nodoi,nourl,noeprint]{revtex4-2}
\usepackage{amsmath,amssymb,physics,bm,siunitx}
\usepackage{xcolor,graphicx,soul}
\usepackage[colorlinks=true,citecolor=blue,urlcolor=blue,linkcolor=black]{hyperref}
\usepackage{subfiles}
\usepackage{setspace}
\usepackage{dcolumn}
\usepackage{diagbox}

\DeclareMathOperator*{\cov}{cov}

\begin{document}

\title{Distributions and correlation properties of offshore wind speeds\\ and wind speed increments}

\author{So-Kumneth Sim}
\email{ssim@uos.de}
\affiliation{Universit\"at Osnabr\"uck, Fachbereich Mathematik/Informatik/Physik, Institut f\"ur Physik,
Barbarastra{\ss}e 7, D-49076 Osnabr\"uck, Germany}

\author{Philipp Maass}
\email{maass@uos.de}
\affiliation{Universit\"{a}t Osnabr\"uck, Fachbereich Mathematik/Informatik/Physik, Institut f\"ur Physik,
Barbarastra{\ss}e 7, D-49076 Osnabr\"uck, Germany}

\author{H. Eduardo Roman}
\email{hector.roman@unimib.it}
\affiliation{Department of Physics, University of Milano-Bicocca,
Piazza della Scienza 3, 20126 Milano, Italy}

\date{July 17, 2024, revised October 18, 2024}

\begin{abstract}
We determine distributions and correlation properties of offshore wind speeds and wind speed increments 
by analyzing wind data sampled with a resolution of one second for 20 months at different heights
above sea level in the North Sea. Distributions of horizontal wind speeds can be fitted to Weibull distributions with
shape and scale parameters varying weakly with the vertical height separation. 
Kullback-Leibler divergences between distributions at different heights change with the squared logarithm of the height ratio.
Cross-correlations between time derivatives of wind speeds 
are long-term anticorrelated, and the even parts of their correlation functions satisfy sum rules.
Distributions of horizontal wind speed increments change from a tent-like shape to a Gaussian with rising increment lag.
A surprising peak occurs in the left tail of the increment distributions for lags in a range $10-200\,\si{km}$ after
applying the Taylor's hypothesis locally to transform time lags into distances. The peak
is decisive in order to obtain an expected and observed linear scaling of third-order structure functions with distance.
This suggests that it is an intrinsic feature of atmospheric turbulence.
\end{abstract}

\maketitle

\section{Introduction}
\label{sec:introduction}
Detailed investigations of wind features are important for testing and further developing theories of atmospheric turbulence. Due to
ongoing improvements of measurement techniques \citep{Suomi/Vihma:2018, vanRamshorst/etal:2020} and 
increasing amount of available data, new insights are obtained by analyzing statistical properties of wind speeds over 
large time and length scales covering many orders of magnitude \citep{Larsen/etal:2016, Sim/etal:2023}. 
Better knowledge of wind properties is of high current interest also for applications, as, for example, for 
harvesting wind energy \citep{Veers/etal:2019}, or for controlling pollutant dispersion in urban areas 
\citep{Leelossy/etal:2014, Tan/Ni:2022}. This includes improved forecasting of wind speeds 
\citep{Costa/etal:2008, Santhosh/etal:2020, Hanifi/etal:2020, Tawn/Browell:2022} and a better understanding 
of wake effects \citep{Sanderse/etal:2011, Archer/etal:2018, Nash/etal:2021, Houck:2022}. 
In addition, spatial correlations on scales of several hundred meters are relevant for mechanical stabilities 
and performances of wind turbines with steadily growing size \citep{Porte-Agel/etal:2020, Bosnjakovic/etal:2022, Houck:2022}.

Stationary distributions of wind speeds $u$ are commonly described by the Weibull distribution, 
where its shape and scale parameters vary with location and season \citep{Johnson:1998, Lun/Lam:2000, Shu/Jesson:2021}.
Distributions of wind speed increments $\Delta_\tau u=u(t)-u(t-\tau)$ in a time interval $\tau$
change in shape strongly with $\tau$. For small lags $\tau$, they decay exponentially for both negative and
positive $\Delta_\tau u$, corresponding to a tent-like form in a linear-log representation 
\citep{Castaing/etal:1990, Boettcher/etal:2007, Milan/etal:2010}. This implies a nonlinear dependence of their moments 
with respect to the moment order, a behavior already considered by Kolmogorov in turbulent flows \citep{Kolmogorov:1962}
and often referred to as intermittency \citep{Mollo:1973, Lohse/Grossmann:1993, Benzi/Vulpiani:2022, Johnson/Wilczek:2024}. 
It is commonly believed that the nonlinear variation of the moments with their order is universal. Different
suggestions have been made to describe it \citep{Kolmogorov:1962, She/Leveque:1994, Boettcher/etal:2007}.
The intermittent features of wind speeds are transferred to wind power feed-in to electricity grids, on the level 
of both a single wind turbine and an entire wind farm \citep{Milan/etal:2010}.

As the lag $\tau$ increases, the shape of increment distributions changes towards that of a Gaussian,
with the kurtosis becoming three when $\tau$ exceeds one or several days. It was suggested \citep{Castaing/etal:1990} 
that the probability density of wind speed increments conditioned on the mean wind speed is given by a linear weighting of 
Gaussian distributions with log-normally distributed standard deviations. By weighting with the Weibull distribution, the unconditioned 
probability density of wind speeds is obtained. To account for the shape variation of the increment distribution, the parameters 
of the log-normal distribution are considered to depend on $\tau$ \citep{Castaing/etal:1990}.

In a complementary approach, commonly referred to as extended self-similarity (ESS) method \citep{Benzi/etal:1993}, the 
moments of absolute values of wind velocity increments are considered \citep{Grossmann/etal:1997}. This approach is based 
on an exact result from the theory of three-dimensional (3D) isotropic turbulence for the third-order moment of spatial 
velocity increments (third-order structure function) and the assumption that this moment is approximately equal to
the third-order moment of absolute values of spatial velocity increments. It was pointed out that a lack of 3D isotropy in the 
turbulent flow, as it is present for near-surface wind flows, narrows the applicability of the ESS method \citep{Amati/etal:1997}. 
Nevertheless, studies of measured and model-generated time series of wind speeds showed that the ESS method is useful, 
with the remarkable result that the scaling of moments with their order is close to that found in isotropic 3D turbulence 
\citep{Vindel/etal:2008, Kiliyanpilakkil/Basu:2016}. In a further application of the ESS method to wind speeds measured at 
different sites in the Netherlands, a universal intermittent behavior was reported \citep{Baile/Muzy:2010}. Applications of the 
ESS method to other geological and geophysical phenomena revealed similar results as for wind \citep{Nikora/Goring:2001}. 
In an earlier study, downward cascades of turbulent atmospheric flows were shown to be reflected in scaling features 
of spatial rainfall and river flow distributions \citep{Gupta/Waymire:1990, Schleiss/etal:2011, Acuna/etal:2020, Leth/etal:2021, Wilby/etal:2023}.
Sufficient conditions for the applicability of the ESS methods to dynamical systems in general were derived in \citep{Yakhot:2001}.

On land, correlation functions of wind speeds show pronounced daily oscillations \citep{Brett/Tuller:1991},
and their properties depend on the local topography \citep{Thomann/Barfield:1988, Perez/etal:2004}.
They are influenced, for example, by buildings \citep{Eliasson/etal:2006} and mountains \citep{Ralph/etal:1997}.
For offshore wind, daily oscillations are not significant in 
correlation functions or spectral analysis \citep{Larsen/etal:2016}, and the fluctuation properties exhibit 
generic statistical features. In a recent study of power spectra $S(f)$, 
distinct frequency regimes were identified displaying characteristic behaviors in agreement
with predictions from theories of atmospheric turbulence \citep{Sim/etal:2023}. These frequency regimes correspond to time regimes, 
which in turn are reflected in spatial domains when applying the Taylor's hypothesis \citep{Taylor:1938}. 

\begin{figure}[t!]
\begin{center}
\includegraphics[width=0.6\columnwidth]{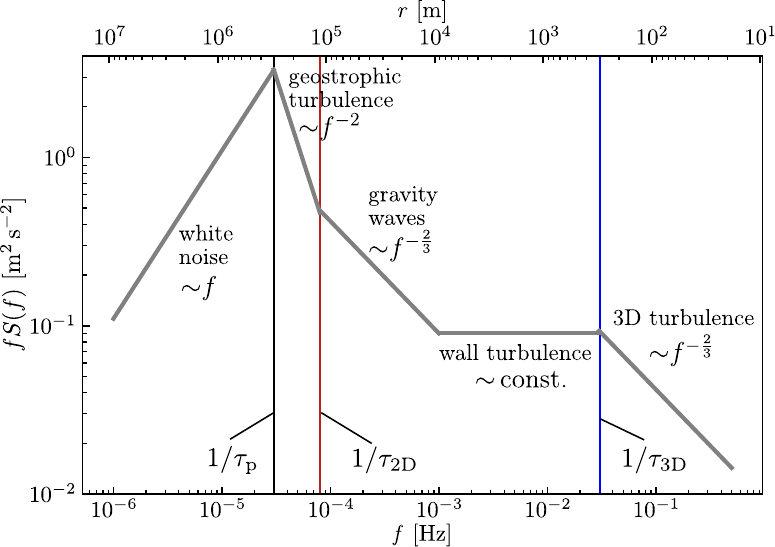}   
\end{center}
\caption{Schematic sketch of the weighted power spectrum $f\,S(f)$ of offshore wind speed fluctuations. 
The upper $x$-axis shows distances corresponding to the frequencies $f$ when using the Taylor's hypothesis,
i.e.\ that wind speeds passing a point at a time difference $\tau$ correspond to wind speeds at a
distance given by the mean wind speed multiplied by $\tau$. For real data, see Ref.~\citep{Sim/etal:2023}.}
\label{fig:illustration-power-spectrum} 
\end{figure}

The distinct frequency regimes can be well identified by looking at the frequency-weighted spectrum, $f\,S(f)$, 
which is presented schematically in Fig.~\ref{fig:illustration-power-spectrum} based on results obtained at the 
measurement height $h=90$\,m \citep{Sim/etal:2023}.
At a frequency $1/\tau_{\rm p}\sim 3\times10^{-5}$\,Hz, 
a peak appears which originates from the motion of low and high-pressure areas in the atmosphere with typical spatial extent of a few 
thousand kilometers. This peak is also seen in spatial wind speed correlations obtained from aircraft measurements 
\citep{Nastrom/Gage:1983, Nastrom/etal:1984}.

%--------------------------------------------------------------------------------------------------
%--------------------------------------------------------------------------------------------------

For frequencies below $1/\tau_{\rm p}$, the frequency-weighted power spectrum $f S$ increases linearly with $f$, corresponding to a constant in the power spectrum $S(f)$. 
Because the analysis for the lowest frequencies was carried out by analyzing
ten-minute averages of wind speeds sampled over 15~years, we are confident that the constant 
$S(f)$ is not due to temporal aliasing effects. 
Spatial aliasing effects resulting from velocity components orthogonal to the mean flow, however,
cannot be ruled out as a possible cause \citep{Pope:2000, Karban/etal:2022}. 
Alternatively, the constant low-frequency spectrum can reflect white noise behavior due to uncorrelated wind speed fluctuations on time scales larger than $\tau_{\rm p}$.  We believe that this is indeed the case.
Further studies would be desirable to clarify this point.

For frequencies above $1/\tau_{\rm p}$, different physical mechanisms lead to correlations between wind speed fluctuations. They govern the spectral behavior 
in distinct regimes: below a frequency $1/\tau_{\rm 2D}\simeq 8\times10^{-5}$\,Hz, $f S\sim f^{-2}$, due to an enstrophy 
cascade in two-dimensional geostrophic turbulence \citep{Charney:1971}. Above $1/\tau_{\rm 2D}$, a regime is dominated by three-dimensional 
turbulence induced by gravity waves, where $f S\sim f^{-2/3}$ occurs \citep{Xie/Buehler:2018, Poblet/etal:2023}. At higher frequencies, this regime is followed
by one referred to as wall turbulence regime, $f S\sim$\,const., where the air flow is strongly affected by the Earth's surface \citep{Calaf/etal:2013, Drobinski/etal:2004}. 
Depending on the height $h$ of the wind speed measurement above the sea level, the wall turbulence regime terminates at a frequency 
$1/\tau_{\rm 3D}\sim \bar v/h$, where $\bar v$ is the mean wind speed. For instance, at $h=100$\,m, we have 
$1/\tau_{\rm 3D}\sim 3\times10^{-2}$\,Hz. For frequencies above $1/\tau_{\rm 3D}$, the spectrum is reflecting 
isotropic 3D turbulence in accordance with Kolmogorov's celebrated law, $S(f)\sim f^{-5/3}$ \citep{Kolmogorov:1941a}, i.e.\ $f S\sim f^{-2/3}$. 

Regarding offshore wind data, previous studies have addressed cross-correla\-tions between equal-time horizontal wind speeds at 
different laterally separated points on the Earth's surface. The cross-correlations were found to decay exponentially for distances 
in the range 4-600\,km, and to stay approximately constant beyond 600\,km \citep{Mehrens/etal:2016}. In the range 2-12\,km, 
the coherence function also decays exponentially with distance \citep{Vincent/etal:2013}.

Measurements performed at a station located at the western tip of the island Fr{\o}ya in Norway provided information
about cross-correlations between horizontal wind speeds in both horizontal and vertical directions to the Earth's surface 
\citep{Bardal/Saetran:2016}. For the horizontal direction, the cross-correlation coefficient between equal-time horizontal 
wind velocities was found to increase with height for velocity components parallel and perpendicular to the mean wind flow. 
For the vertical direction, the cross-correlation turned out to decay slower with the separation distance if the mean wind 
speed becomes higher.

In this work, we investigate distribution functions of wind speeds and their increments,
on different times scales, as well as time correlation functions between the respective quantities.
In particular, we analyze cross-correlations between 
points separated by a vertical distance $\Delta h$. 

%--------------------------------------------------------------------------------------------------
\begin{figure}[b!]
\begin{center}
\includegraphics[width=0.5\columnwidth]{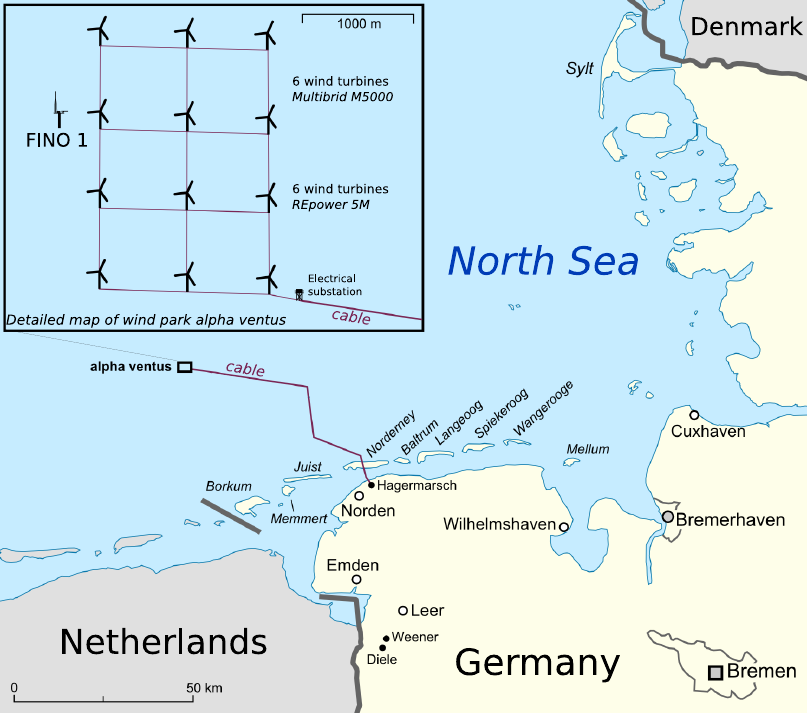}   
\end{center}
\caption{Location of the FINO1 measurement mast in the North Sea. 
It is situated near the wind farm ``alpha ventus'', approximately 45~km north of the island of Borkum.
The figure is a modified version of the map
\href{https://de.wikipedia.org/wiki/Offshore-Windpark_alpha_ventus\#/media/Datei:Windpark_alpha_ventus_Lagekarte.png}{Wind park alpha ventus} 
from Lencer (CC BY-SA 3.0).}
\label{fig:FINO-platform} 
\end{figure}
%--------------------------------------------------------------------------------------------------

\section{Wind speed data from FINO1}
\label{sec:data}    
Wind speeds were measured at the FINO1 platform in the North Sea, which is located about $45\,\si{km}$ north of the island Borkum
\footnote{{FINO1} project supported by the German Government through BMWi and PTJ. For further 
details on the data sampling and instrumentation, see \texttt{https://www.fino1.de/en}.}, 
see Fig.~\ref{fig:FINO-platform}.  The wind speed data were sampled by three-cup anemometers over
20\,months, from September 2015 to April 2017, for eight different heights $h$ between $30\,\si{m}$ and $100\,\si{m}$.  The 
time resolution is $\Delta t=1\,\si{s}$, yielding time series with $N\cong 5\times10^7$ wind speeds for each height. Because a 
lightning rod can influence the measurement at $h=100$\,m, we exclude the corresponding data from our analysis.

The time series of wind speeds contain missing values, with the fraction of these values being less than 1\% for all heights.
For the quantities analyzed in this work, we have checked that no special treatment was necessary for interpolating 
missing values. When calculating any averaged quantity, results obtained for different interpolation schemes did not differ 
significantly from those obtained when omitting the missing values.

Due to the lack of simultaneous data for wind directions and wind speeds, we cannot perform
an analysis separately for different wind directions. The prevailing wind direction at the FINO1 platform is from the west, corresponding 
to an onshore flow direction \citep{Karagali/etal:2013}.  We thus expect that our findings are not significantly affected
by wind flows from land. We would also like to note that variations in stratification \citep{Deusebio/etal:2014, Segalini/Arnqvist:2015, Yassin/Griffies:2022, Platis/etal:2022}
can impact our analysis.

\section{Theoretical methods and analysis}
\label{sec:methods}
\subsection{Distributions of horizontal wind speeds at different heights}
\label{subsec:methods_similarity_distributions}
PDFs of wind speeds are commonly described by the Weibull distribution 
\citep{Tuller/Brett:1984, Chang:2011, Mohammadi/etal:2016, Sim/etal:2019, Shu/Jesson:2021}
\begin{equation}
\label{eq:PWeibull}
\psi(u) = \frac{k}{\lambda}\left(\frac{u}{\lambda}\right)^{k-1}e^{-\left(\frac{u}{\lambda}\right)^k}
\end{equation}
with $\lambda$ the scale and $k$ the shape parameter. 
These parameters are estimated by
a nonlinear fit of Eq.~\eqref{eq:PWeibull} to the PDF obtained from the wind speed data, where we have used
an equidistant binning of the 
wind speeds with bin sizes in the range $0.42-0.52\,\si{m/s}$.

According to the Weibull distribution, the relative standard deviation of wind speeds, also referred to as turbulence intensity,
is independent of the scale parameter:
\begin{equation}
\frac{\sigma_u}{\bar u}=\sqrt{\frac{\Gamma^2(1+1/k)}{\Gamma(1+2/k)}-1}\,.
\label{eq:turbulene-intensity}
\end{equation}
Here, $\Gamma(.)$ is the Gamma function, and $\bar u$ and $\sigma_u$ are the mean and standard deviation of $\psi(u)$.

To quantify the difference between wind speed distributions $\psi(u;h_1)$ and $\psi(u;h_2)$ at two different heights $h_1$ and $h_2$, 
we make use of the Kullback-Leibler
divergence \citep{Kullback:1997,Teixeira/etal:2019}
\begin{align}
&\mathcal{D}[\psi(u;h_1),\psi(u;h_2)]= \int_{0}^{\infty}\mathrm{d}u\,{\psi(u;h_1)\log\frac{\psi(u;h_1)}{\psi(u;h_2)}}
\label{eq:KL}\\
&\hspace{1em}= \log\left(\frac{k_1}{k_2}\frac{\lambda_2^{k_2}}{\lambda_1^{k_1}}\right)
+(k_1-k_2)\left(\log\lambda_1 - \frac{\gamma}{k_1}\right)
+ \left[\left(\frac{\lambda_1}{\lambda_2}\right)^{k_2}\Gamma\left(\frac{k_2}{k_1} + 1\right) - 1 \right]\nonumber\,,
\end{align}
where $\gamma$ is the Euler-Mascheroni constant, and $k_j$ and $\lambda_j$ are the shape and scale parameters of
$\psi(u)$ at height $h_j$. Specifically, we consider the symmetrized Kullback-Leibler (or Jeffrey) divergence, 
\begin{align}
D_{\rm KL}(h_1,h_2)&=\frac{1}{2}\left(\mathcal{D}[\psi(u;h_1),\psi(u;h_2)]+\mathcal{D}[\psi(u;h_2),\psi(u;h_1)]\right)
\nonumber\\
&=\frac{1}{2}\log\left[ \left(\frac{\lambda_1}{\lambda_2}\right)^{k_1}\left(\frac{\lambda_2}{\lambda_1}\right)^{k_2}\right]+
\frac{\gamma}{2}\left(\frac{k_1}{k_2}+\frac{k_2}{k_1} \right)-\gamma-1
\nonumber\\
&\hspace{1em}{}+\frac{1}{2}\left(\frac{\lambda_1}{\lambda_2}\right)^{k_2} \Gamma\left(\frac{k_2}{k_1}+1\right) +
\frac{1}{2}\left(\frac{\lambda_2}{\lambda_1}\right)^{k_1} \Gamma\left(\frac{k_1}{k_2}+1\right)\,.
\label{eq:KL-symmetrized}
\end{align}

%-------------------------------------------------------------------------------------------------------------------------------------------------------
\subsection{Temporal cross-correlations of wind speeds and wind speed increments at different heights}
\label{subsec:methods_cross-correlations}
Temporal cross-correlations between fluctuations of 
a quantity $X(t,h_1)$ at height $h_1$ and $X(t,h_2)$ at height $h_2$ 
are investigated based on the normalized cross-correlation function
\begin{equation}
C_X(t;h_1,h_2) = \frac{\left\langle[X(t',h_1)-\bar X(h_1)][X(t'+t,h_2)-\bar X(h_2)]\right\rangle_{t'}}{\sigma_X(h_1)\sigma_X(h_2)}
\label{eq:CCF}
\end{equation}
where $\langle\ldots\rangle_{t'}$ denotes an average over $t'$, and
$\bar X(h)=\langle X(t,h)\rangle_t$ and $\sigma_X(h)=[\langle X^2(t,h)\rangle_t-\langle X(t,h)\rangle_t^2]^{1/2}$ are the mean and standard deviation 
of $X$ at height $h$, respectively. For $t=0$, these cross-correlation functions equal Pearson correlation coefficients
between equal-time fluctuations of $X$ at different heights. For $h_1=h_2=h$, $C_X(t;h,h)$ is the normalized autocorrelation function of $X$ at height $h$. It holds
$|C_X(t;h,h)|\le C_X(0;h,h)=1$ and $|C_X(t;h_1,h_2)|\le1$ for $h_1\ne h_2$.

We study cross-correlations of both the wind speeds [$X(t,h)=u(t,h)$] and of their time derivatives $\partial_t u(t,h)\cong u(t+\Delta t,h)-u(t,h)$ 
[$X(t,h)=u(t+\Delta t,h)-u(t,h)$, $\Delta t=1\si{s}$], i.e.\ horizontal wind accelerations.
In power spectra of the offshore wind speeds, we have not seen any signatures of diurnal and annual variations \citep{Sim/etal:2023}.
Diurnal variations in near-surface marine wind are mainly observed when it flows in offshore direction 
\citep{Lapworth:2005}, while the wind at the FINO1 platform flows primarily from the west
in onshore direction \citep{Karagali/etal:2013}.
Annual variations show up as sharp peak in the power spectrum of wind speeds when analyzing 10-minutes averaged wind speed data sampled over 20~months, 
but it was found also that these do not significantly affect the correlation properties of the wind speed on time scales of 20~months.

We thus regard the wind speed as a stationary process in our
correlation analysis. When decomposing the correlations functions into their even and odd parts,
i.e.\  $C_X(t;h_1,h_2)=C_X^+(t;h_1,h_2)+C_X^-(t;h_1,h_2)$ with 
\begin{equation}
C_X^\pm(t;h_1,h_2)=\frac{1}{2}\left[C_X(t;h_1,h_2)\pm C_X(-t;h_1,h_2)\right]\,,
\end{equation}
the even parts
should fulfill the relation
\begin{equation}
C^+_u(t;h_1,h_2)=C^+_u(0;h_1,h_2)-
\frac{\sigma_{\partial_t u}(h_1)\sigma_{\partial_t u}(h_2)}{\sigma_u(h_1)\sigma_u(h_2)}\int\limits_0^t\dd t' (t-t')\,
C^+_{\partial_t u(t';h_1,h_2)}\,,
\label{eq:CuCdotu-relation}
\end{equation}
and $C^+_{\partial_t u}(t;h_1,h_2)$ the two sum rules
\begin{align}
\int_0^\infty\dd t\, C^+_{\partial_t u(t;h_1,h_2)}&=0
\label{eq:sum-rule-1}\,,\\
\int_0^\infty\dd t\, C^+_{\partial_t u}(t;h_1,h_2)\,t&
=-\frac{\cov[u(t,h_1)u(t,h_2)]}{\sigma_{\partial_t u}(h_1)\sigma_{\partial_t u}(h_2)}\,.
\label{eq:sum-rule-2}
\end{align}
Here, $\cov[u(t,h_1),u(t,h_2)]=\langle [u(t,h_1)-\bar u(h_1)][u(t,h_2)-\bar u(h_2)]\rangle$ is the covariance between 
equal-time wind speeds at heights $h_1$ and $h_2$, which becomes the variance $\sigma_u^2(h)$ for $h_1=h_2=h$.
Note that the auto-correlation functions are even functions of time, $C_X^-(t;h,h)=0$.

Equations~\eqref{eq:CuCdotu-relation}-\eqref{eq:sum-rule-2} can be derived by using
\begin{equation}
u(t,h)-u(0,h)=\int\limits_0^t\dd t'\, \frac{\partial u(t',h)}{\partial t'}\,,
\end{equation}
and by taking into account that  $\partial_t u(t,h)$ is a stationary process. It thus holds
\begin{align}
&\langle [u(t,h_1)\!-\!u(0,h_1)][u(t,h_2)\!-\!u(0,h_2)]\rangle
=\int\limits_0^t\dd t_1 \int\limits_0^t\dd t_2 \left\langle \frac{\partial u(t_1,h_1)}{\partial t_1}\frac{\partial u(t_2,h_2)}{\partial t_2}\right\rangle\nonumber\\
&\hspace{3em}=\sigma_{\partial_t u}(h_1)\sigma_{\partial_t u}(h_2)\int\limits_0^t\dd t_1 \int\limits_0^t\dd t_2\,C_{\partial_t u}(t_1-t_2;h_1,h_2)
\label{eq:Delta_tu-C-relation}\\
&\hspace{3em}=2\sigma_{\partial_t u}(h_1)\sigma_{\partial_t u}(h_2)\left[t\int\limits_0^t\dd t'\, C^+_{\partial_t u}(t';h_1,h_2)
-\int\limits_0^t\dd t'\, t' C^+_{\partial_t u}(t';h_1,h_2)
\right]\,.
\nonumber
\end{align}
The simplification of  the double integral of $C_{\partial_t u}(t_1-t_2;h_1,h_2)$ over $t_1$ and $t_2$ 
to the single integrals in the last line is possible due to the fact that the integrand is a function of $(t_1-t_2)$ only.

Defining $\delta u(t,h)=u(t,h)-\bar u(h)$, 
the left-hand side of Eq.~\eqref{eq:Delta_tu-C-relation} can be rewritten as
\begin{align}
&\langle [u(t,h_1)-u(0,h_1)][u(t,h_2)-u(0,h_2)]\rangle\nonumber\\[0.5ex]
&\hspace{3em}=\langle [\delta u(t,h_1)-\delta u(0,h_1)][\delta u(t,h_2)-\delta u(0,h_2)]\rangle\nonumber\\[0.5ex]
&\hspace{3em}=2\langle \delta u(0,h_1)\delta u(0,h_2)\rangle
-\left[\langle \delta u(t,h_1)\delta u(0,h_2)\rangle+\langle \delta u(0,h_1)\delta u(t,h_2)\rangle\right]\nonumber\\[0.5ex]
&\hspace{3em}=2\cov[u(t,h_1),u(t,h_2)]-2\sigma_u(h_1)\sigma_u(h_2)\, C^+_u(t;h_1, h_2)\,.
\end{align}
Equating this with the right-hand side of Eq.~\eqref{eq:Delta_tu-C-relation} yields Eq.~\eqref{eq:CuCdotu-relation}.

When considering the limit $t\to\infty$ in Eq.~\eqref{eq:Delta_tu-C-relation}, the left-hand side becomes two times the covariance 
$\cov[u(t,h_1),u(t,h_2)]$. Accordingly, the
first integral in the square bracket in Eq.~\eqref{eq:Delta_tu-C-relation} must vanish, which yields Eq.~\eqref{eq:sum-rule-1}, and the remaining second integral
in Eq.~\eqref{eq:Delta_tu-C-relation} must satisfy Eq.~\eqref{eq:sum-rule-2}.

To quantify the correlation between wind speed increments
\begin{equation}
\Delta_\tau u(t,h)=u(t,h)-u(t-\tau,h)\,,
\end{equation}
for time lags $\tau>1\,\si{s}$ at different heights, we calculate the Pearson correlation coefficient 
\begin{equation}
\rho(\tau;h_1,h_2)=C_{\Delta_\tau u}(0;h_1,h_2)
=\frac{\langle\Delta_\tau u(t,h_1)\Delta_\tau u(t,h_2)\rangle}{\sqrt{\langle\Delta_\tau u^2(t,h_1)\rangle\langle\Delta_\tau u^2(t,h_2)\rangle}}\,.
\label{eq:rho-coefficient}
\end{equation}

%---------------------------------------------------------------------------------------------------------------------------------------------------------------
\subsection{Detrended fluctuation analysis of differences between wind speeds in vertical direction}
\label{subsec:methods_DFA_vertical_wind_speed_increments}   
To study how differences 
\begin{equation}
\Delta u_{h_1,h_2}(t) = u(t,h_1) - u(t,h_2)
\label{eq:difvel} 
\end{equation}
between horizontal wind speeds at different heights $h_1$ and $h_2$ fluctuate in time, we apply 
the detrended fluctuation analysis (DFA) \citep{Koscielny-Bunde/etal:1998}.
This is a valuable method in particular in the presence of long-range power-law correlations.

In the DFA, the $\Delta u_{h_1,h_2}(t)$ are considered as positions of a random walk. 
A sub- or superlinear power-law scaling of 
its mean squared displacement with time signals long-range power-law decays of
the correlation function $C(t;\,\Delta u_{h_1,h_2},\Delta u_{h_1,h_2})$.
To uncover such correlations clearly, the DFA includes a detrending due to local biases.

Specifically, we define a profile of the detrended random walk in a time window $t$ by
\begin{equation}
R_{h_1,h_2}(t')=\sum_{t^{\prime\prime}=1}^{t'} \Delta u_{h_1,h_2}(t^{\prime\prime})
-\frac{t'}{t}\sum_{t^{\prime\prime}=1}^t \Delta u_{h_1,h_2}(t^{\prime\prime})\,,\hspace{1em}
1\le t'\le t\,,
\label{eq:profile} 
\end{equation}
where the second term removes a linear bias in the time window. Note that $R_{h_1,h_2}(0)=R_{h_1,h_2}(t)=0$.
The time series of total duration $T$ is divided into $M=\lfloor T/t\rfloor$ time windows of size $t$, 
where $\lfloor x\rfloor$ is the largest integer smaller than $x$.
The mean of the profile in the $m$th window is
\begin{equation}
\bar R_{h_1,h_2}(m)= \frac{1}{t}\sum_{t'=1}^t R_{h_1,h_2}[(m-1)t+t']\,,\hspace{1em}m=1,\ldots,M\,.
\label{eq:meanRWprof}
\end{equation}
It is a measure of the mean of detrended wind speed differences in the $m$th window on time scale $t$. Differently
speaking, it is the $m$th position of the detrended random walk, when it is coarse-grained on time scale $t$.
The variance 
\begin{equation}
F_{h_1,h_2}(t)=F_{h_2,h_1}(t)=\frac{1}{M-1}\sum_{m=1}^{M-1} [\bar R_{h_1,h_2}(m+1)-\bar R_{h_1,h_2}(m)]^2
\label{eq:fluctuationfunction}
\end{equation}
of the displacements $\bar R_{h_1,h_2}(m+1)-\bar R_{h_1,h_2}(m)$ of this random walk quantifies
the strength of fluctuations of $\Delta u_{h_1,h_2}(t)$ on time scale $t$. It is called the fluctuation function in the DFA.

For power-law behavior 
\begin{equation}
F_{h_1,h_2}(t) \sim t^H\,,
\label{eq:Hurstexponent}
\end{equation}
at long times with $0<H<1$, $H$ is commonly referred to as the Hurst exponent.
Values  $1/2< H<1$ correspond to a long-range power-law decay
$\sim t^{2H-2}$ of the autocorrelation function $\langle\partial_t\Delta u_{h_1,h_2}(t')\partial_t\Delta u_{h_1,h_2}(t'+t)\rangle_{t'}$.

%-----------------------------------------------------------------------------------------------------------------------------------------
\subsection{Distribution functions of horizontal wind speed increments at different Taylor distances}
\label{subsec:updf}
For analyzing distributions of horizontal wind speed increments at a constant height, we make use of the Taylor's hypothesis.
Applying it requires sufficiently low turbulence intensities $\sigma_u/\bar u$. For our data, $\sigma_u/\bar u=0.52$, which
is at the limit where the hypothesis was found to break down \citep{Willis/Deardorff:1976}.
Indeed, in a previous analysis of the offshore wind data \citep{Sim/etal:2023}
we have seen that Taylor's hypothesis does not give 
reliable results when applying it with the mean wind speed calculated from the whole time series. 

However, reliable results were obtained if the hypothesis was applied
locally: the scaling of structure functions agrees well with theoretical predictions for atmospheric turbulence, and with 
results obtained from an analysis of wind fields measured by aircraft \citep{Cho/Lindborg:2001}.
Here we also apply the Taylor's hypothesis locally
to convert data from the time to the spatial domain.

In this method, we first calculate the average wind speed $\bar u_{t,t+\tau}$ in the interval $[t,t+\tau[$. The time interval is then converted into a distance
$r_{t,t+\tau}=\bar u_{t,t+\tau}\,\tau$, which we refer to as the Taylor distance for time lag $\tau$ at time $t$.
For each pair of times $(t, t+\tau)$, 
we calculate the distance $r=r_{t,t+\tau}$, and the corresponding wind speed increment $\Delta_r u\equiv u_t-u_{t+\tau}$. 
The $\Delta_r u$ are then grouped with respect to equally spaced Taylor distances $r$ on a 
logarithmic scale. The $j$th group contains all wind speed increments for $r$ lying in the interval
$[r_j^-,r_{j+1}^-[$, $j=0,1,\ldots, j_{\rm max}$, where $r_j^-=r_010^{j/10}$ with $r_0=1$\,m\,.
The value of $r$ assigned to the $j$th group is the geometric mean $r_j=(r_j^-r_{j+1}^-)^{1/2}$, 
corresponding to the arithmetic mean of logarithms, $\ln r_j=(\ln r_j^-+\ln r_{j+1}^-)/2$. 
For each $r_j$, the wind speed increments $\Delta_r u$ 
are sampled in small bins $\Delta_r u$ to obtain the PDFs $p_r(\Delta_r u)$.

%-------------------------------------------------------------------------------------------------------------------------------------------------------
\subsection{Scaling behavior of moments of horizontal wind speed increment distributions}
\label{subsec:methods_ESS}
Characteristic turbulence features can be
identified through the scaling behavior of structure functions, which are 
moments of $p_r(\Delta_r u)$ of order $q$:
\begin{equation}
D_q(r)=\left\langle (\Delta_r u)^q\right\rangle\,.
\label{eq:structf}
\end{equation}
The third-order structure function $D_3$ is expected to follow known 
behaviors in certain distance or time regimes. In the regime of 3D isotropic
turbulence, $D_3$ is negative, proportional to the energy dissipation rate $\epsilon$ and
grows linearly with $r$ \citep{Kolmogorov:1941c, Taylor/etal:2003}:
\begin{equation}
D_3(r)=-\frac{4}{5} \epsilon r\,.
\label{eq:D3_3D}
\end{equation}
Here, $\epsilon$ is the energy dissipation rate.
For gravity-wave induced turbulence, Coriolis forces cannot be neglected, which leads to a modified 
amplitude in the linear scaling of the third-order structure function \citep{Lindborg/Cho:2001}:
\begin{equation}
D_3(r)=-2 \epsilon r\,.
\label{eq:D3_GW}
\end{equation}
In quasi-2D geostrophic turbulence, 
$D_3$ is positive and scales as 
\begin{equation}
D_3(r)=\frac{1}{4} \eta r^3\,,
\label{eq:D3_GS}
\end{equation}
where $\eta$ is the enstrophy dissipation rate \citep{Lindborg:1999, Lindborg/Cho:2001, Lindborg:2007, Xie/Buehler:2018, Poblet/etal:2024}.
In the regime of wall turbulence, a simple scaling of $D_3$ with $r$ is not expected but measurements have shown that $D_3$ 
is negative in general  \citep{Sim/etal:2023}.

To examine intermittency in the regime of 3D isotropic turbulence, we consider absolute values of wind speed increments.
The corresponding moments are denoted as $\tilde{D}_q(r)$ to distinguish them from the ones in Eq.~\eqref{eq:structf}:
\begin{equation}
\tilde{D}_q(r)=\left\langle |\Delta_r u|^q\right\rangle\,.
\label{eq:structf_abs}
\end{equation}
As $D_3\sim -r$ in the considered regime, one can expect that $\tilde D_3\sim r$ also, which is indeed obtained.
As for other moments, a self-similar structure of the turbulent field \citep{Kolmogorov:1941a}
would imply a monofractal behavior $\tilde{D}_q\sim \tilde{D}_3^{q/3}\sim r^{q/3}$.
However, due to intermittent behavior, 
\begin{equation}
\tilde{D}_q(r) \sim r^{\zeta(q)}\,,
\label{eq:structf_scaling}
\end{equation}
or
\begin{equation}
\tilde{D}_q(r) \sim \tilde D_3^{\zeta(q)}\,,
\label{eq:structf_ESSanalysis}
\end{equation}
where $\zeta(q)$ is a nonlinear function of $q$ with $\zeta(3)=1$. 
Deviations from the linear dependence $\zeta(q) = q/3$ give insight
into the multifractal scaling due to intermittency. 
Based on a turbulent cascade model with log-normally distributed 
energy transfer rates, Kolmogorov \citep{Kolmogorov:1962} derived
\begin{equation}
\label{eq:zetaq_Kolmogorov}            
\zeta(q) = \frac{q}{3} - \frac{\mu}{18}q(q-3)\,,
\end{equation}
where the intermittency correction $\mu$ lies in the range $0.2-0.5$ \citep{Sreenivasan/Kailasnath:1993, Praskovsky/Oncley:1997}.
Assuming high velocity fluctuations on small spatial 
scales, another nonlinear behavior
\begin{equation}
\label{eq:zetaq_SheLeveque}            
\zeta(q) = \frac{q}{9} + 2\left[1 - \left(\frac{2}{3}\right)^{q/3}\right]\,,
\end{equation}
was proposed by She and Leveque \citep{She/Leveque:1994}. Several other functions $\zeta(q)$
were suggested in the literature \citep{Lovejoy/etal:2001,Yakhot:1998}.
We will determine exponents $\zeta(q)$ by fitting power laws 
to data of $\tilde D_q$ vs.\ $r$ [Eq.~\eqref{eq:structf_scaling}] 
and of $\tilde D_q$ vs.\ $\tilde D_3$ [Eq.~\eqref{eq:structf_ESSanalysis}] in the inertial range of 3D isotropic turbulence.
The fitting to data of $\tilde D_q$ vs.\ $\tilde D_3$ corresponds to the ESS analysis \citep{Benzi/etal:1993}.
Exponents $\zeta(q)$ obtained from the power-law fits are checked against the theoretical
predictions \eqref{eq:zetaq_Kolmogorov} and \eqref{eq:zetaq_SheLeveque}.

\section{Results}
\label{sec:results}
We discuss our results according to the subsectioning in the Theoretical Methods outlined in Sec.~\ref{sec:methods}.

%------------------------------------------------------------------------------------------------------------------------------------------
\subsection{Distributions of horizontal wind speeds at different heights}
\label{sect:KLresults}
\begin{figure}[t!]
\centering
\includegraphics[width=0.6\columnwidth]{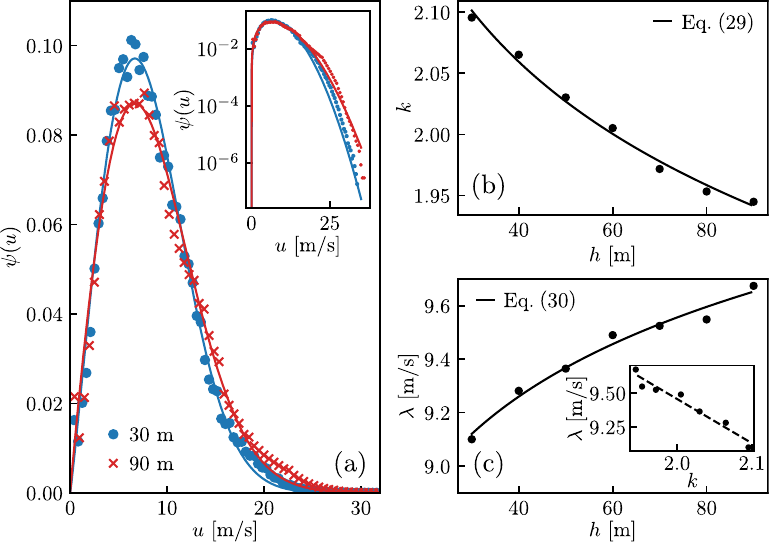}   
\caption{(a) PDFs of wind speeds at heights $30\,\si{m}$ (dots) and $90\,\si{m}$ (crosses) and fits
of Weibull distributions to the data [Eq.~\eqref{eq:PWeibull}] (solid lines). The fitting was carried out by using the Levenberg-Marquardt algorithm.
The inset shows the PDFs together with the fits in the semi-logarithmic representation.
(b) Shape parameters $k$ (circles) and (c) scale parameters $\lambda$ (circles) of the fitted Weibull distributions
as a function of height $h$. The lines in (b) and (c) are least-square fits 
to Eqs.~\eqref{eq:k-h} and \eqref{eq:lambda-h}.
The inset in (c) shows the $\lambda$- in relation to the $k$-values (circles) with
the dashed line being a linear least-square fit to the data, cf.\ Eq.~\eqref{eq:lambda-h}.}
\label{fig:Weibull_k_lambda} 
\end{figure}

PDFs of the wind speeds sampled for the two heights $h=30\,\si{m}$ and  $h=90\,\si{m}$ (symbols) 
are displayed in Fig.~\ref{fig:Weibull_k_lambda}(a) together with fits of the Weibull distribution to the data (solid lines).
For the fitting, we applied the Levenberg-Marquardt algorithm \citep{Fischer/etal:2024}.
The Weibull distribution gives a good description
except in the tail regime where it underestimates the frequency of high horizontal wind speeds $u \gtrsim 18\,\si{m/s}$.
The deviation in the tail regime can be seen in the semilogarithmic representation in the inset in Fig.~\ref{fig:Weibull_k_lambda}(a).
The shape parameter $k$ is close to two, in agreement with earlier studies 
reported in the literature \citep{Chang:2011, CostaRocha/etal:2012, Tizgui/etal:2017, Patidar/etal:2022}, 
see Fig.~\ref{fig:Weibull_k_lambda}(b).
The solid line in Fig.~\ref{fig:Weibull_k_lambda}(b) indicates a 
weak logarithmic dependence of $k$ on $h$,
\begin{equation}
k(h)=k_0-\alpha\ln\left(\frac{h}{h_0}\right)\,,
\label{eq:k-h}
\end{equation}
where we have taken $h_0=30\,\si{m}$ as the reference height, $k_0=k(h_0)\cong2.0036$, and $\alpha\cong 0.15$. 
The shape parameter $\lambda$ decreases approximately linearly with $k$, see the inset of Fig.~\ref{fig:Weibull_k_lambda}(c). This yields
\begin{equation}
\lambda(h)=\lambda_0-\lambda_{\rm s} [k(h)-k_0]=\lambda_0+\alpha\lambda_{\rm s}\ln(h/h_0)
\label{eq:lambda-h}
\end{equation}
for the $h$-dependence, where the coefficients are $\lambda_0\cong9.12\,\si{m/s}$ and $\lambda_{\rm s}\cong3.33\,\si{m/s}$. The solid line
in Fig.~\ref{fig:Weibull_k_lambda}(c) marks the dependence on $h$ according to Eq.~\eqref{eq:lambda-h}.

%------------------------------------------------------------------------------------------------------------------------------------------------------
%------------------------------------------------------------------------------------------------------------------------------------------------------

%-------------------------------------------------------------------------------------------------------------------------------------------------------
\begin{figure}[b!]
\centering
\includegraphics[width=0.6\columnwidth]{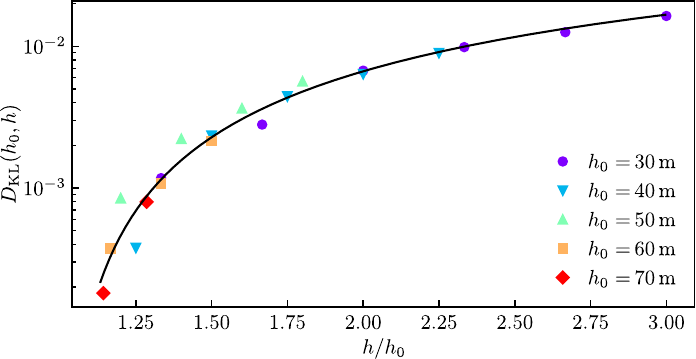}   
\caption{Symmetrized Kullback-Leibler divergences $D_\mathrm{KL}(h_0,h)$ [symbols, see Eq.~\eqref{eq:KL-symmetrized}]
as a function of $h/h_0$. For $k_1,k_2$ and $\lambda_1,\lambda_2$ in Eq.~\eqref{eq:KL-symmetrized}, the parameters 
from the fits of the Weibull distributions in
Figs.~\ref{fig:Weibull_k_lambda}(b) and \ref{fig:Weibull_k_lambda}(c) are used. The solid line gives $D_\mathrm{KL}(h_0,h)$ according to 
Eq.~\eqref{eq:DKL-h}.} 
\label{fig:KLdiv_v} 
\end{figure}
%--------------------------------------------------------------------------------------------------------------------------------------------------------

The mean velocity $\bar u$ and the turbulence intensity $\sigma_u/\bar u$ are related to $k$ and $\lambda$:
\begin{align}
\bar u(h)&=\Gamma\left(1+\tfrac{1}{k}\right)\lambda\cong \frac{\sqrt{\pi}}{2}\left[\lambda_0+\alpha\lambda_{\rm s}\ln\left(\frac{h}{h_0}\right)\right]\,,
\label{eq:baru-h}\\
\frac{\sigma_u}{\bar u}&=\sqrt{\frac{\Gamma\left(1+\tfrac{2}{k}\right)}{\Gamma^2\left(1+\tfrac{1}{k}\right)}-1}\cong
\sqrt{\frac{4}{\pi}-1}\cong0.52\,.
\label{eq:turbulenceintensity-h}
\end{align}
Here, we have evaluated the Gamma functions at $k=2$ in the approximate expressions. In agreement with
earlier findings \citep{Karlsson:1986, Marusic/etal:2013} and theories of wall turbulence \citep{Tennekes:1973, Townsend:1976}, Eq.~\eqref{eq:baru-h} gives a
logarithmic increase of the mean wind speed $\bar u$ with height $h$.

Figure~\ref{fig:KLdiv_v} shows how the dissimilarity between the wind speed PDFs, as quantified by 
the symmetrized Kullback-Leibler 
divergence $D_\mathrm{KL}(h_0, h)$ [see Eq.~\eqref{eq:KL-symmetrized}],
increases with $h/h_0$. 
Using Eqs.~\eqref{eq:k-h} and \eqref{eq:lambda-h}, we can express this dependence.
To this end, $D_\mathrm{KL}(h_0,h)$ in
Eq.~\eqref{eq:KL-symmetrized} is considered as a function of $k$, i.e.\ $k_1$, $\lambda_1$ in Eq.~\eqref{eq:KL-symmetrized} become
$k_0$, $\lambda_0$, and $k_2$, $\lambda_2$ become $k$, $\lambda(k)$, with $\lambda(k)$ the linear function in Eq.~\eqref{eq:lambda-h}. 
Since $k$ varies only weakly with $h$,
we can expand this function $D_\mathrm{KL}$ of $k$ in
a Taylor series around $k_0$. Up to second order, we obtain ($D_\mathrm{KL}=0$ and $\dd D_\mathrm{KL}/\dd k=0$ for $k=k_0$),
\begin{align}
D_{\rm KL}(h_0,h)&=\frac{1}{2}\left[\left(1-\gamma+\frac{\lambda_{\rm s}}{\lambda_0}\,k_0^2\right)^2+\frac{\pi^2}{6}\right]\,\frac{\Delta k^2}{k_0^2}
\nonumber\\
&=\frac{\alpha^2}{2k_0^2}\left[\left(1-\gamma+\frac{\lambda_{\rm s}}{\lambda_0}\,k_0^2\right)^2+\frac{\pi^2}{6}\right]\,\ln^2\left(\frac{h}{h_0}\right)\,.
\label{eq:DKL-h}
\end{align}
In the last step, we have inserted $\Delta k=-\alpha\ln(h/h_0)$ from Eq.~\eqref{eq:k-h}. The solid line in Fig.~\ref{fig:KLdiv_v} demonstrates that $D_{\rm KL}(h_0,h)$
is well described by Eq.~\eqref{eq:DKL-h}.

%---------------------------------------------------------------------------------------------------------------------------------------------------------
\subsection{Temporal cross-correlations of wind speeds and wind speed increments at different heights}
\label{subsec:resuls_cross-correlations}

Kullback-Leibler divergences in the previous section quantify differences between distributions of the wind speeds, 
but do not allow conclusions regarding correlations. In this subsection, we discuss correlation properties of
wind speeds, wind accelerations and wind speed increments.

\begin{figure}[b!]
\centering
\includegraphics[width=0.55\columnwidth]{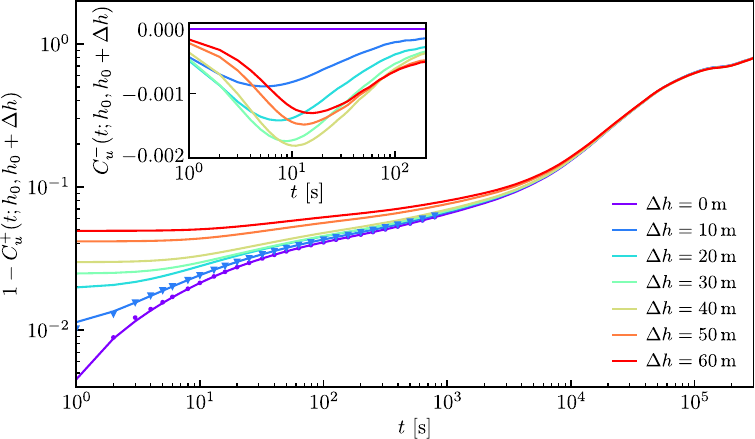}   
\caption{Even parts $C_u^+(t;h_0,h_0+\Delta h)$ of cross-correlation functions between wind speeds [solid lines, see Eq.~\eqref{eq:CCF}] 
for various height separations $\Delta h$ from the
reference height $h_0=30\,\si{m}$. For better visibility of the relative deviations between the curves for different $\Delta h$, we have plotted $1-C_u^+(t;h_0,h_0+\Delta h)$. The symbols for $\Delta h=0\,\si{m}$ and $\Delta h=10\,\si{m}$ are obtained from Eq.~\eqref{eq:CuCdotu-relation}, 
where the integral in this equation was calculated numerically. The inset shows the odd parts $C_u^-(t;h_0,h_0+\Delta h)$ of the cross-correlation functions.}
\label{fig:CCFu} 
\end{figure}

Figure~\ref{fig:CCFu} shows the even parts $C_u^+(t;h_0,h_0+\Delta h)$ (main graph) and odd parts $C_u^-(t;h_0,h_0+\Delta h)$ (inset) of the correlation functions $C_u(t;h_0,h_0+\Delta h)$ for a reference height 
$h_0=30\,\si{m}$ and different vertical separations $\Delta h$. The odd parts for $\Delta h\ne 0$ are negative and run through a minimum at values around $t\sim 10\,\si{s}$. They are for all times much smaller than the even parts ($|C_u^-(t;h_0,h_0+\Delta h)/C_u^+(t;h_0,h_0+\Delta h)|<2\times 10^{-3}$), so that 
$C_u(t;h_0,h_0+\Delta h)\simeq C_u^+(t;h_0,h_0+\Delta h)$. For the even parts, we have plotted $1-C_u^+(t;h_0,h_0+\Delta h)$ rather than $C_u^+(t;h_0,h_0+\Delta h)$ in the double-logarithmic representation to show that
relative deviations between the cross-correlation functions for different $\Delta h$ are strong at small times.
For large times $t\gtrsim 10^4\,\si{s}$, they become negligible. 
The $C_u^+(t;h_0,h_0+\Delta h)$ decay
very slowly, i.e.\ the curves  in Fig.~\ref{fig:CCFu} slowly approach one for large $t$. The curve for $\Delta h=0$ 
is proportional to the second-order structure function, 
$1-C_u^+(t;h_0,h_0)=D_2(t,h_0)/2\sigma_u^2(h_0)$, where
$D_2(t,h_0)=\langle u(t',h_0)-u(t'+t)\rangle_{t'}$.

%--------------------------------------------------------------------------------------------------------------------------------------------------
%---------------------------------------------------------------------------------------------------------------------------------------------------

%--------------------------------------------------------------------------------------------------------------------------
\begin{figure}[t!]
\centering
\includegraphics[width=0.75\columnwidth]{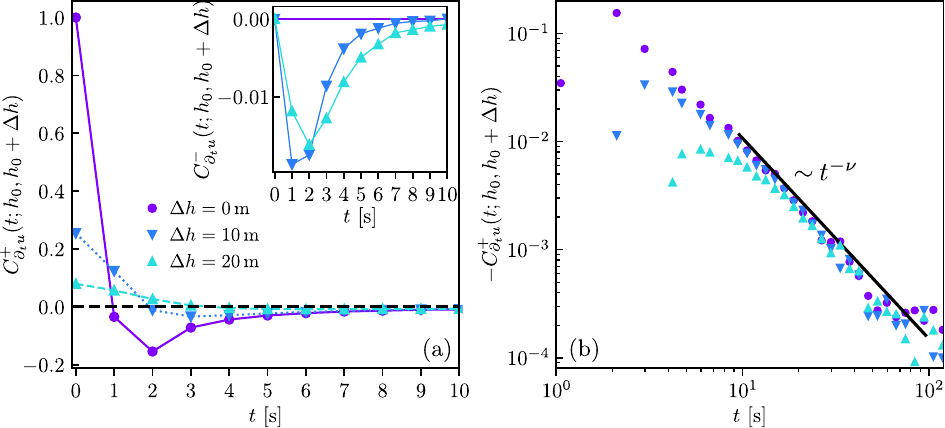}   
\caption{(a) Even parts $C_{\partial_t u}^+(t;h_0,h_0+\Delta h)$ of cross-correlation functions
between time derivatives of wind speeds (symbols)
at vertical separations $\Delta h=0$, 10, $20\,\si{m}$ from the reference height $h_0=30\,\si{m}$.
Lines are connecting the data as a guide for the eye.
The inset shows the odd parts of the cross-correlation functions, 
including the autocorrelation analogous to Fig.~\ref{fig:CCFu}.
In (b), the even parts are shown in a double-logarithmic representation
up to larger times. The solid line is a least-square fit of a power law to the autocorrelation data in the range
$10\,\si{s}\le t\le 100\,\si{s}$, see Eq.~\eqref{eq:Cdtu-power-law}.}
\label{fig:CCFdu}  
\end{figure}
%--------------------------------------------------------------------------------------------------------------------------

Figure~\ref{fig:CCFdu}(a) shows the even parts $C_{\partial_t u}^+(t;h_0,h_0+\Delta h)$ (main graph) and odd parts $C_{\partial_t u}^-(t;h_0,h_0+\Delta h)$ (inset) of the correlation functions $C_{\partial_t u}(t;h_0,h_0+\Delta h)$ for 
the same reference height $h_0=30\,\si{m}$ as in Fig.~\ref{fig:CCFu}. The odd parts are negative for all $t$ and
run through a minimum at short times for $\Delta h\ne0$. 
For the even part, we checked  the relation~\eqref{eq:CuCdotu-relation} between
$C^+_{\partial_t u}(t;h_0,h_0+\Delta h)$ and $C^+_u(t;h_0,h_0+\Delta h)$ and
that the $C^+_{\partial_t u}(t;h_0,h_0+\Delta h)$ satisfy the 
sum rule~\eqref{eq:sum-rule-1}. The validity of Eq.~\eqref{eq:CuCdotu-relation} is demonstrated
by the symbols in Fig.~\ref{fig:CCFu}, which were determined by numerical calculations 
of the integral in Eq.~\eqref{eq:CuCdotu-relation}.
The confirmation of the relation~\eqref{eq:CuCdotu-relation} up to times $t\gtrsim 10^3\,\si{s}$ 
suggests that it holds true for all times and hence also
the second sum rule \eqref{eq:sum-rule-2}. We could not verify it directly since
the $C^+_{\partial_t u}(t;h_0,h_0+\Delta h)$ do not have sufficient numerical accuracy for $t\gtrsim 10^3\,\si{s}$.

The autocorrelation function $C_{\partial_t u}^+(t;h_0,h_0)$ in Fig.~\ref{fig:CCFdu}(a)
rapidly decreases from its value one at $t=0$ to a negative minimal value and thereafter approaches zero smoothly. 
The functional form of $C_{\partial_t u}^+(t;h_0,h_0+\Delta h)$
for $\Delta h>0$ is similar, but $|C_{\partial_t u}^+(t;h_0,h_0+\Delta h)|$ of the cross-correlations is reduced for small times
$t\lesssim 7\,\si{s}$ compared to the autocorrelations. For larger times, relative deviations between the $C_{\partial_t u}^+(t;h_0,h_0+\Delta h)$ for different
$\Delta h$ are insignificant, as Fig.~\ref{fig:CCFdu}(b) illustrates. The double-logarithmic plot suggests that the
decay in the time regime $10\,\si{s}\lesssim t\lesssim 10^2\,\si{s}$ can be described by a power law,
\begin{equation}
C_{\partial_t u}^+(t;h_0,h_0+\Delta h)\sim - A\,t^{-\nu}\,,
\label{eq:Cdtu-power-law}
\end{equation}
where $A\cong 0.7$ and $\nu\cong1.9$. 

\begin{figure}[t!]
\centering
\includegraphics[width=0.7\columnwidth]{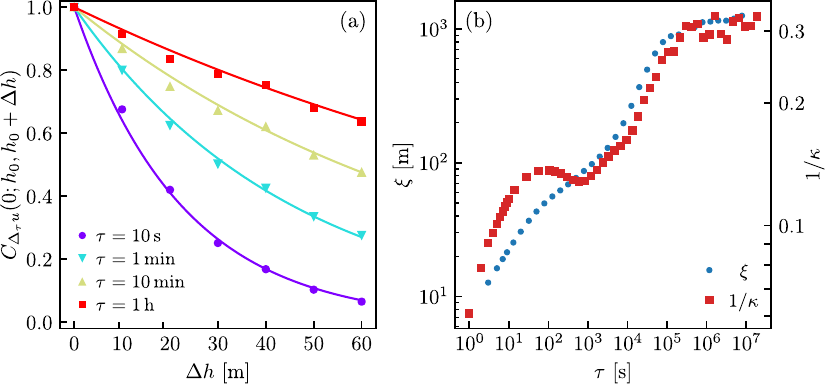}   
\caption{({\bf a}) Pearson correlation coefficients between horizontal wind speed increments [symbols, see Eq.~\eqref{eq:rho-coefficient}]  
as a function of height difference $\Delta h$ for various increment time lags $\tau$ and reference height $h_0=30\,\si{m}$.
Solid lines are least-square fits to Eq.~\eqref{eq:corrlength}.
({\bf b}) Correlation lengths $\xi$ obtained from the fits  (circles) and inverse kurtosis $1/\kappa$ of the increment distributions (squares) 
as a function of $\tau$.}
\label{fig:CCFdu0} 
\end{figure}

Figure~\ref{fig:CCFdu0}(a) shows the Pearson correlation coefficients $\rho(\tau;h_0,h_0+\Delta h)$ defined
in Eq.~\eqref{eq:rho-coefficient} as a function of $\Delta h$ for various time lags $\tau$ and the same reference height $h_0=30\,\si{m}$ as in Figs.~\ref{fig:CCFu} and
\ref{fig:CCFdu}.
The coefficients quantify how wind speed increments $\Delta_\tau u$ at vertical
separation $\Delta h$ are correlated. In all cases, the data can be well fitted by an exponential 
decay, 
\begin{equation}
\rho(\tau;h_0,h_0+\Delta h)=\exp[-\frac{\Delta h}{\xi(\tau)}]\,,
\label{eq:corrlength}
\end{equation}
where $\xi(\tau)$ is a $\tau$-dependent correlation length. 

The correlation length $\xi$ increases with $\tau$, see Fig.~\ref{fig:CCFdu0}(b). For $\tau$-scales corresponding to the
regime of 3D isotropic turbulence (see Introduction), $\xi(\tau)$ is of the order of $10\,\si{m}$, i.e.\ smaller than the measurement 
heights $h$. For $\tau\gtrsim10^5\,\si{s}$, $\xi$ approaches $1\,\si{km}$, which is
comparable to the planetary boundary layer height \citep{Zhang/etal:2013}. 
This height is the upper limit, where vertical convection has a significant influence on the spatial distribution of horizontal wind speeds.

%--------------------------------------------------------------------------------------------------------------------------
%------------------------------------------------------------------------------------------------------------------------- 

Interestingly, the variation of $\xi$ with $\tau$ is similar to that of $1/\kappa(\tau,h)$ for fixed $h$,
where $\kappa(\tau,h)$
is the kurtosis of the wind speed increment distribution at height $h$, $\kappa(\tau,h)=\left\langle \Delta_\tau u^4(t,h)\right\rangle_t/\left\langle \Delta_\tau u^2(t,h)\right\rangle_t^2$. 
We have included the plot of $1/\kappa(\tau,h_0)$ as a function of $\tau$ in Fig.~\ref{fig:CCFdu0}(b)
for the reference height $h_0=30\,\si{m}$.
For other $h$, the corresponding curve would look almost the same, i.e.\ there
are no significant changes in the dependence of $1/\kappa(\tau,h)$ on $\tau$ when varying $h$.
The similarity in the behavior of $\xi(\tau)$ and $1/\kappa(\tau,h)$ can be interpreted as follows:
when $\kappa(\tau,h)$ is large, the probability for the occurrence of very large wind speed increments is high. Large wind speed increments
disturb coherent turbulent structures and hence spatial correlations. Accordingly, one can expect $\xi$ to become 
smaller for larger $\kappa$ in agreement with the behavior seen in Fig.~\ref{fig:CCFdu0}(b).

%------------------------------------------------------------------------------------------------------------------------------------------------------
\begin{figure}[b!]
\centering
\includegraphics[width=0.8\columnwidth]{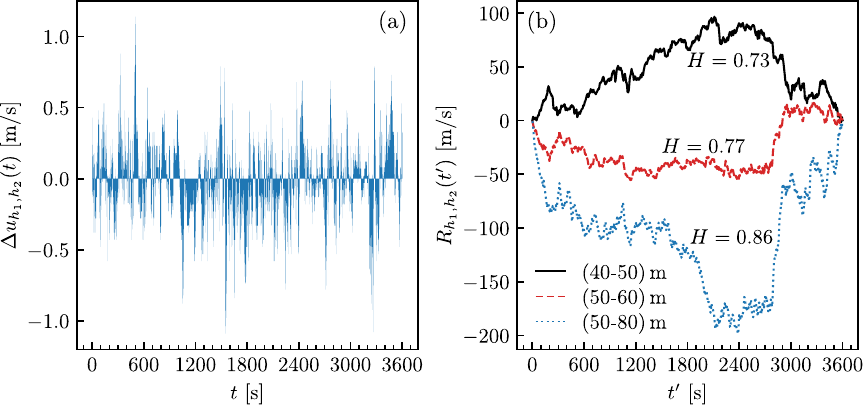}
\caption{(a) One-hour section of the time series of wind speed differences 
$\Delta u_{h_1,h_2}(t) = u(t,h_1) - u(t,h_2)$ at heights $h_1=40\,\si{m}$ and $h_2=50\,\si{m}$.
(b) Profile $R_{h_1,h_2}(t)$ of the detrended random walk in the one-hour time window generated from the data in (a) according to Eq.~\eqref{eq:profile} (black). Two further profiles in the same one-hour time window are shown for heights $h_1=50\,\si{m}$, $h_2=60\,\si{m}$ (red) and $h_1=50\,\si{m}$, $h_2=80\,\si{m}$ (blue).}
\label{fig:Delta-u_profile}
\end{figure}
%-----------------------------------------------------------------------------------------------------------------------------------------------------------------------------

%-------------------------------------------------------------------------------------------------------------------------------------------------------------
\subsection{Detrended fluctuation analysis of differences between wind speeds in vertical direction}
\label{subsec:results_DFA_vertical_wind_speed_increments}
A representative example of a 
one-hour section of the time series of wind speed differences 
$\Delta u_{h_1,h_2}(t) = u(t,h_1) - u(t,h_2)$ for heights $h_1=40\,\si{m}$ and $h_2=50\,\si{m}$ is shown in Fig.~\ref{fig:Delta-u_profile}(a). 
The patchy structure of 
this series suggests the presence of long-time correlations \citep{Koscielny-Bunde/etal:1998}. 
In Fig.~\ref{fig:Delta-u_profile}(b), we display profiles $R_{h_1,h_2}(t)$ of the detrended random walk 
calculated from Eq.~\eqref{eq:profile} for the time window in Fig.~\ref{fig:Delta-u_profile}(a). The black line corresponds to the data in
Fig.~\ref{fig:Delta-u_profile}(a) [$h_1=40\,\si{m}$, $h_2=50\,\si{m}$]. Two further profiles for heights
$h_1=50\,\si{m}$, $h_2=60\,\si{m}$ (red) and $h_1=50\,\si{m}$, $h_2=80\,\si{m}$ (blue) are shown also.

\begin{figure}[t!]
\centering
\includegraphics[width=0.6\columnwidth]{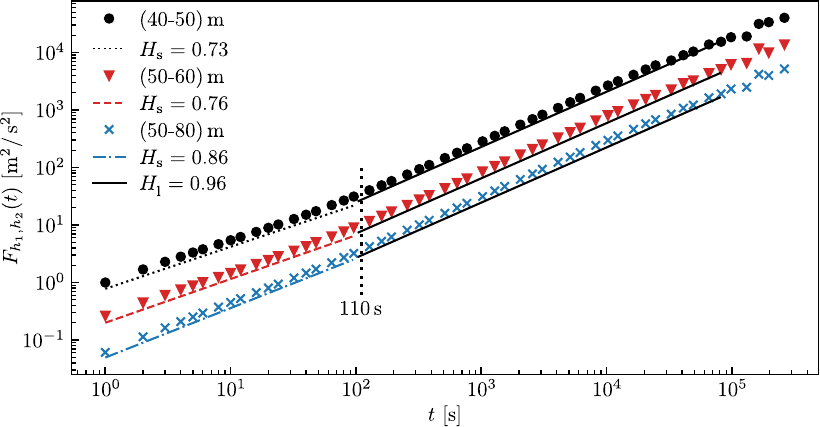}
\caption{Fluctuation functions $F_{h_1,h_2}(t)$ for $h_1=50\,\si{m}$ and $h_2=40\,\si{m}$, 60\,m, and 80\,m.
The solid and dashed lines indicate power-law behavior in short- and long-time regimes with Hurst exponents 
$H_{\rm s}$ and $H_{\rm l}$, respectively. 
$H_{\rm s}$ varies in the range 0.73-0.92 for different heights $h_1$, $h_2$, see Table~\ref{tab:Hs}, 
while $H_{\rm l}\simeq 0.96$ is approximately independent of the heights. 
The vertical lines mark the crossover time $t_\times\simeq110\si{s}$ between the two time regimes. 
It does not show a significant dependence on the height.}
\label{fig:F(t)}
\end{figure}

Fluctuation functions $F_{h_1,h_2}(t)$ calculated from the profiles, see Eq.~\eqref{eq:fluctuationfunction}, 
are depicted in Fig.~\ref{fig:F(t)} for $h_1=50\,\si{m}$, and
$h_2=40\,\si{m}$, 60\,m, and 80\,m. One can identify two regimes, 
separated by a crossover time $t_\times\simeq 110\,\si{s}$, where $F(t)$ shows different power-law behavior.
In the long-time regime $t>t_\times$, fluctuation functions scale as $F(t)\sim t^{H_{\rm l}}$, with a Hurst exponent $H_{\rm l}\simeq 0.96$
showing no significant variation with heights $h_1$, $h_2$. 
In the short-time regime, $t<t_\times$, $F(t)\sim t^{H_{\rm s}}$, where the Hurst exponent $H_{\rm s}$ is smaller than 
$H_{\rm l}$ and depends on $h_1$ and $h_2$. The values vary in the range $0.73\le H_S\le 0.92$, see Table~\,\ref{tab:Hs}.

% -----------------------------------------------------------------------------------------------------------------
\begin{table}[b!]
\caption{\label{tab:Hs} Hurst exponents $H_{\rm s}$ characterizing the power-law increase of the fluctuation functions $F_{h_1,h_2}(t)$  [see Eq.~\eqref{eq:Hurstexponent}].  The $H_{\rm s}$  are the slopes of the solid lines in the double-logarithmic plot of the DFA fluctuation 
functions $F_{h_1,h_2}(t)$ in Fig.~\ref{fig:F(t)} for times $t\lesssim 110\,\si{s}$.
Due to the symmetry $F_{h_1,h_2}(t)=F_{h_2,h_1}(t)$, Hurst exponents are the same for $h_1$ and $h_2$ interchanges. 
In the table, we give the values for $h_1<h_2$.}
\begin{tabular}{| c | c | c | c | c | c | c | c |} 
\hline 
\backslashbox{$h_1$}{$h_2$}
&   30\,m   &    40\,m    &   50\,m    &   60\,m   &   70\,m   &   80\,m    &   90\,m\\ 
\hline 
30\,m      &    -     &  0.73   &  0.80  &  0.85  &  0.89  &  0.91  &  0.92\\
40\,m      &    -     &    -      &  0.73  &  0.81  &  0.86  &  0.89  &  0.90\\
50\,m      &   -      &    -      &    -     &  0.76  &  0.83  &  0.86  &  0.88\\
60\,m      &    -     &    -      &    -     &    -     &  0.75  &  0.82  &  0.85\\
70\,m      &    -     &     -     &    -     &    -     &    -     &  0.75  &  0.81\\
80\,m      &    -     &     -     &    -     &    -     &    -     &    -     &  0.74\\
\hline
\end{tabular}
\end{table}
% -----------------------------------------------------------------------------------------------------------------

%-----------------------------------------------------------------------------------------------------------------------------------------------------------------------------
%-----------------------------------------------------------------------------------------------------------------------------------------------------------------------------

The long-range correlations reflected by the large Hurst exponent $H_{\rm l}\simeq0.96$ demonstrate a high 
persistence of fluctuations of wind speed differences at different heights. The short-time exponents $H_{\rm s}$
listed in Table~\ref{tab:Hs} indicate a weakly increasing persistence for larger height differences. This result is
unexpected and should be considered with care as the relation between power-law behavior in $F_{h_1,h_2}(t)$
and corresponding power-law decays of $\langle\partial_t\Delta u_{h_1,h_2}(t')\partial_t\Delta u_{h_1,h_2}(t'+t)\rangle_{t'}$
is less robust than for the long-time asymptotic behavior.

%-----------------------------------------------------------------------------------------------------------------------------------------------------------------------------
\subsection{Distributions of horizontal wind speed increments on different length scales}
\begin{figure}[t!]
\begin{center}
\includegraphics[width=0.9\columnwidth]{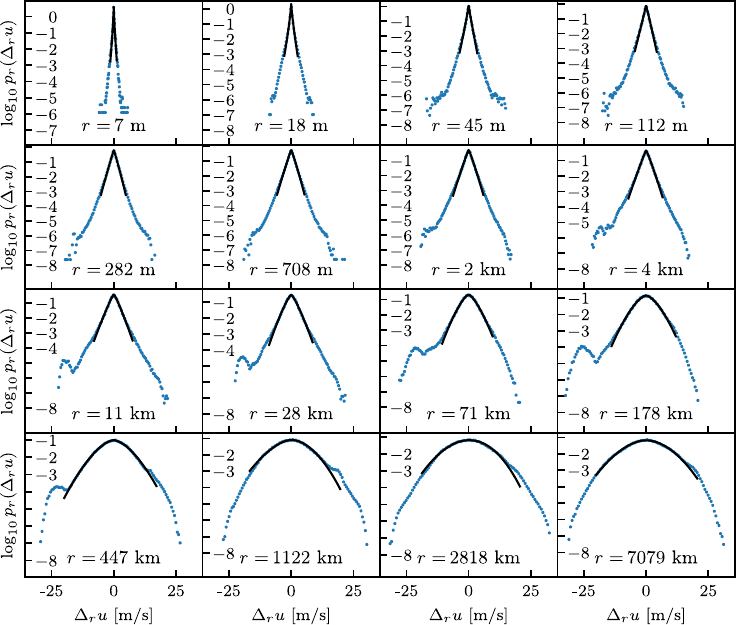}   
\end{center}
\caption{PDFs $p_r(\Delta_r u)$ of horizontal wind speed increments $\Delta_r u$ for 
16 different Taylor distances $r$ at height $h=90\,\si{m}$.
The black lines indicate least-square fits of the data to Eq.~\eqref{eq:core-pdf} in the core part $[\Delta_r u^-,\Delta_r u^+]$, where
$\Delta_r u^-$ and $\Delta_r u^-$ are the first and last $g$-quantiles for $g=10^3$, see Eq.~\eqref{eq:core-part}.}
\label{fig:pdf_u}
\vspace*{2ex}
\end{figure}

PDFs $p_r(\Delta_r u)$ of horizontal wind speed increments at 16 different Taylor distances $r$ 
[see Sec.~\ref{subsec:updf}] are displayed in Fig.~\ref{fig:pdf_u} for the
height $h=90$\,m. At small $r\lesssim 100\,\si{m}$, the distributions have a tent-like shape, 
which reflects the intermittent behavior of 3D isotropic turbulence \citep{Tabeling/etal:1996, Boettcher/etal:2007}.
In contrast to a Gaussian distribution, strong changes of wind speeds over distances $r$, or corresponding time intervals 
$\tau\simeq r/\bar u$, are much more frequent, which has important consequences for wind park engineering and other 
wind-sensitive applications \citep{Peinke/etal:2004, Ren/etal:2018}.

%-----------------------------------------------------------------------------------------------------------------------------------------------------
%-----------------------------------------------------------------------------------------------------------------------------------------------------

For larger $r$, the tent-like shape continues to be present in a core interval $\Delta_r u\in[\Delta_r u^-,\Delta_r u^+]$ 
around the maximum at $\Delta_r u=0$. We define this core part by
\begin{equation}
\int\limits_{-\infty}^{\Delta_r u^-} \dd x\, p_r(x)=\int\limits_{\Delta_r u^+}^{\infty} \dd x\, p_r(x)=\frac{1}{g}\,,
\label{eq:core-part}
\end{equation}
i.e.\ $\Delta_r u^-$ and $\Delta_r u^+$ are the first and last $g$-quantile. 
Taking $g=1000$, we can
fit the PDFs in $[\Delta_r u^-,\Delta_r u^+]$
by skewed stretched exponential functions,
\begin{equation}
\label{eq:core-pdf}
p_r(\Delta_r u)\propto
\left\{\hspace*{-0.25em}\begin{array}{ll}
\exp{\left(-|\Delta_r u/s_-|^{\alpha_-}\right)}\,, & \Delta_r u\le0\,,\\[1ex]
\exp{\left(-|\Delta_r u/s_+|^{\alpha_+}\right)}\,, & \Delta_r u>0\,,
\end{array}\right.
\end{equation}
where the scale parameters $s_\pm$ quantify the widths
and the shape parameters $\alpha_\pm$ describe 
the decays for negative and positive $\Delta_r u$, respectively. Corresponding least-square fits
are shown as solid lines in Fig.~\ref{fig:pdf_u}.

\begin{figure}[t!]
\begin{center}
\includegraphics[width=0.75\columnwidth]{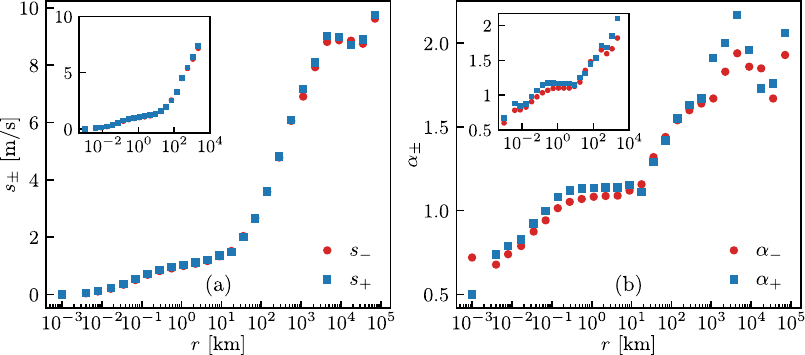}   
\end{center}
\caption{({\bf a}) Scale parameters $s_\pm$ and ({\bf b}) shape parameters $\alpha_\pm$ specifying the 
core part of the distribution of horizontal wind speed increments according to Eq.~\eqref{eq:core-pdf} 
(see also solid lines in Fig.~\ref{fig:pdf_u}). Data in the main figure parts are for height $h=90$\,m and 
in the insets for $h=30$\,m.}
\label{fig:updf-parameters} 
\end{figure}

In Fig.~\ref{fig:updf-parameters}, we display the fit parameters for $h=90\,\si{m}$, and in the insets the parameters for $h=30\,\si{m}$.
For both $h=30$\,m and $90$\,m, the parameters are nearly the same, demonstrating that the core part of the wind 
speed increment PDFs is almost independent of $h$. The scale parameters $s_\pm$ in Fig.~\ref{fig:updf-parameters}(a) 
increase with $r$, reflecting the broadening of the PDFs in Fig.~\ref{fig:pdf_u}. 

This broadening can also be quantified by the second-order structure functions $D_2(r)$, see Eq.~\eqref{eq:structf}.  In fact,
by assuming Eq.~\eqref{eq:core-pdf} to describe the PDFs for all $\Delta_r u$, one finds
\begin{equation}
D_2=a^2(s_-^2+b^2s_+^2)\,,
\end{equation}
where
\begin{equation*}
b^2=\frac{\alpha_+\Gamma(2/\alpha_-)}{\alpha_-\Gamma(2/\alpha_+)}\simeq 1,
\end{equation*}
and
\begin{equation*}
a^2=\frac{b^2\,\frac{\Gamma(1/\alpha_+)}{\alpha_+}+ \frac{\Gamma(1/\alpha_-)}{\alpha_-}}
{b^2\,\frac{\Gamma(3/\alpha_+)}{\alpha_+}+ \frac{\Gamma(3/\alpha_-)}{\alpha_-}}\,.
\end{equation*}
The shape parameters $\alpha_\pm$ in Fig.~\ref{fig:updf-parameters}(b) increase from values $\alpha_\pm < 1$ for small $r$, 
to values $\alpha_\pm\simeq 2$ for large $r$. Accordingly, deviations from a Gaussian distribution become less pronounced for 
larger distances $r$, as it can be seen also by the change of shape of the core part from tent-like to parabolic in Fig.~\ref{fig:pdf_u}.

In addition to the features describing the core part of the distributions, bumps occur in both the left and right tail of the PDFs in 
Fig.~\ref{fig:pdf_u}, see, for example, the results for $r=4$\,km and $r=1122$\,km. The bumps in the right 
tail appear for $r\gtrsim 400$\,km and are small. The bumps in the left tail become more pronounced with increasing 
$r$ and develop into a peak around $\Delta_r u\simeq -25$\,m/s, see the PDFs in the range $10$\,km $<r<450$\,km. 
When the Taylor distance changes from $r=447$\,km to $r=1122$\,km, the peak suddenly disappears.

%--------------------------------------------------------------------------------------------------------------------------------------------------------
%----------------------------------------------------------------------------------------------------------------------------------------------------------------

\begin{figure}[t!]
\centering
\includegraphics[width=0.65\columnwidth]{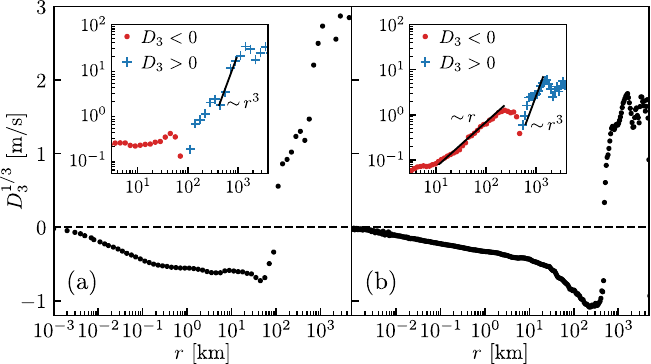}   
\caption{Third-order structure function at height $h=90\,\si{m}$ as a function of the Taylor distance $r$ for
(a) the core part of the distribution of horizontal wind speed increments [PDFs given by Eqs.~\eqref{eq:core-pdf}], and
(b) the full PDFs. To make changes more clearly visible, the third root $D_3^{1/3}$ is plotted. The insets 
show the third-structure function in a double-logarithmic representation, where red circles indicate $-D_3$ (sign conversion 
for negative $D_3$) and blue crosses $+D_3$ (no sign conversion). In the interval $500-1000\,\si{km}$, the solid lines in the 
insets have slope three according to the scaling behavior $D_3(r)\sim r^3$ predicted by theory for geostrophic turbulence, 
see Eq.~\eqref{eq:D3_GS} and Fig.~\ref{fig:illustration-power-spectrum}.
In the inset  in (b), a second solid line in the interval $10-200\,\si{km}$ has slope one according to the scaling, $D_3(r)\sim -r$ predicted 
for 3D turbulence induced by gravity waves, see Eq.~\eqref{eq:D3_GW}.}
\label{fig:D3} 
\end{figure}

Interestingly, the second peak at negative $\Delta_r u$ has a strong impact on the change of sign
in the third-order structure function $D_3(r)$ and its scaling behavior in the regime 10-200\,km. This is demonstrated 
in Fig.~\ref{fig:D3}, where in (a), $D_3(r)$ is plotted for the core part, assuming Eq.~\eqref{eq:core-pdf} 
to provide a description for all $\Delta_r u$, while in Fig.~\ref{fig:D3}(b), $D_3$ was calculated for all data including the 
bumps in the tails of the PDF. In both Figs.~\ref{fig:D3}(a) and (b), $D_3$ changes sign, but in Fig.~\ref{fig:D3}(a) this 
change occurs at a significantly smaller $r\simeq 100$\,km, compared to $r\simeq 500$\,km in Fig.~\ref{fig:D3}(b).
In view of previous results reported in the literature \citep{Cho/Lindborg:2001, Gkioulekas/Tung:2006, Callies/etal:2014, Poblet/etal:2023, Sim/etal:2023}, 
the value $r\simeq 500$\,km is more reliable, indicating that the tail features of the PDF have a decisive influence. 

%----------------------------------------------------------------------------------------------------------------------------------------------------------------
%------------------------------------------------------------------------------------------------------------------------------------------------------------------

This finding is further corroborated when analyzing the scaling behavior of $D_3$, where $D_3=-2\epsilon r$ is expected for 
the regime of 3D turbulence induced by gravity waves [Eq.~\eqref{eq:D3_GW}], and $D_3=\eta r^3/4$ for the regime of 
geostrophic turbulence [Eq.~\eqref{eq:D3_GS}], see also Fig.~\ref{fig:illustration-power-spectrum}. 
In the insets of Figs.~\ref{fig:D3}(a) and (b), 
$D_3(r)$ is plotted in double-logarithmic representation.
For the 3D turbulence regime induced by gravity waves,
the expected scaling $D_3\sim -r$ is clearly visible in the inset of Fig.~\ref{fig:D3}(b), while it cannot be seen in the inset of Fig.~\ref{fig:D3}(a).
The solid lines with slopes three indicate a scaling $D_3\sim r^3$ as it is expected in the regime of geostrophic turbulence. 
The $D_3$ from our analysis show a strong rise in the respective $r$ regime, which may reflect such scaling behavior. 
Due to the small range of the regime and the scatter of the $D_3$ data in it, however, 
one cannot speak about an agreement with theoretical predictions.

%-------------------------------------------------------------------------------------------------------------------------------------------------
\begin{figure}[t!]
\begin{center}
\includegraphics[width=0.7\columnwidth]{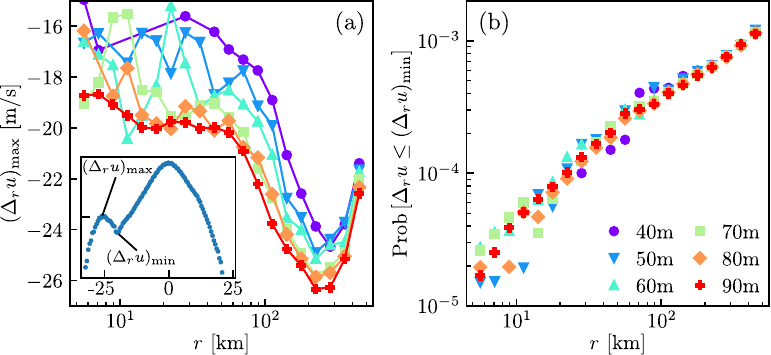}   
\end{center}
\caption{(a) Horizontal wind speed increments $(\Delta_r u)_{\rm max}$, where $p_r(\Delta _r u)$ shows a local maximum in its left tail,
see the example in the inset, where $p_r(\Delta _r u)$ is shown for $h=90\,\si{m}$ and $r=178\,\si{km}$; for further examples, see
Fig.~\ref{fig:pdf_u}. Also indicated in the inset is the increment $(\Delta_r u)_{\rm min}$, where $p_r(\Delta _r u)$ has a minimum.
(b) Probability of finding a speed increment $\Delta_r u < (\Delta_r u)_{\rm min}$.
Data in (a) and (b) are shown for six different heights. Assignment of symbols and color coding are given in the legend in (b).}
\label{fig:argmax_p(u)} 
\end{figure}
%-------------------------------------------------------------------------------------------------------------------------------------------------

We denote the wind speed increments, where $p_r(\Delta u)$ exhibits a local minimum and maximum in its left tail as $(\Delta_r u)_{\rm max}$ 
and $(\Delta_r u)_{\rm min}$, see the inset of Fig.~\ref{fig:argmax_p(u)}(a).
How $(\Delta_r u)_{\rm max}$ varies with the Taylor distance $r$ for various heights $h$ is shown in Fig.~\ref{fig:argmax_p(u)}(a). 
With increasing $h$, $(\Delta_r u)_{\rm max}$ decreases,
i.e.\ the position of the second peak shifts further into the left tail part of $p_r(\Delta _r u)$.
For all $h$, $(\Delta_r u)_{\rm max}$ becomes minimal at a Taylor distance of about $250$\,km. Interestingly, this value
coincides with the minimum of the third-order structure function $D_3$, see Fig.~\ref{fig:D3}(b). 

The data suggest that increments $\Delta u_r<(\Delta_r u)_{\rm min}$ are resulting from a different physical mechanism than increments  $\Delta u_r>(\Delta_r u)_{\rm min}$ distributed according to the core part of $p_r(u)$. 
The probability $\mathrm{Prob}\left[\Delta_r u \le (\Delta_r u)_{\mathrm{min}}\right]$ for increments $\Delta u_r<(\Delta_r u)_{\rm min}$ to occur,
is depicted in Fig.~\ref{fig:argmax_p(u)}(b).
It increases approximately linearly with $r$.

To summarize, our analysis indicates that the peak in the left tail of distributions of horizontal wind speed increments 
is decisive to obtain the scaling features of $D_3(r)$ according to gravity-wave induced turbulence, Eq.~\eqref{eq:D3_GW}.
The distance, at which this peak appears at largest negative speed increments $(\Delta_r u)_{\rm max}$,
agrees with the minimum of $D_3(r)$. 
We have checked also that the peak appears robustly when 
bootstrapping the data by determining increment PDFs for ten separate chunks of two months.
These findings suggest
that the peak in the left tail is not an artifact but an 
important, so-far unexplored feature. That it was not reported yet, could be due to the lack of sufficient 
amount of data for resolving it in the tail part of PDFs. 

%--------------------------------------------------------------------------------------------------
\begin{figure}[b!]
\begin{center}
\includegraphics[width=0.77\columnwidth]{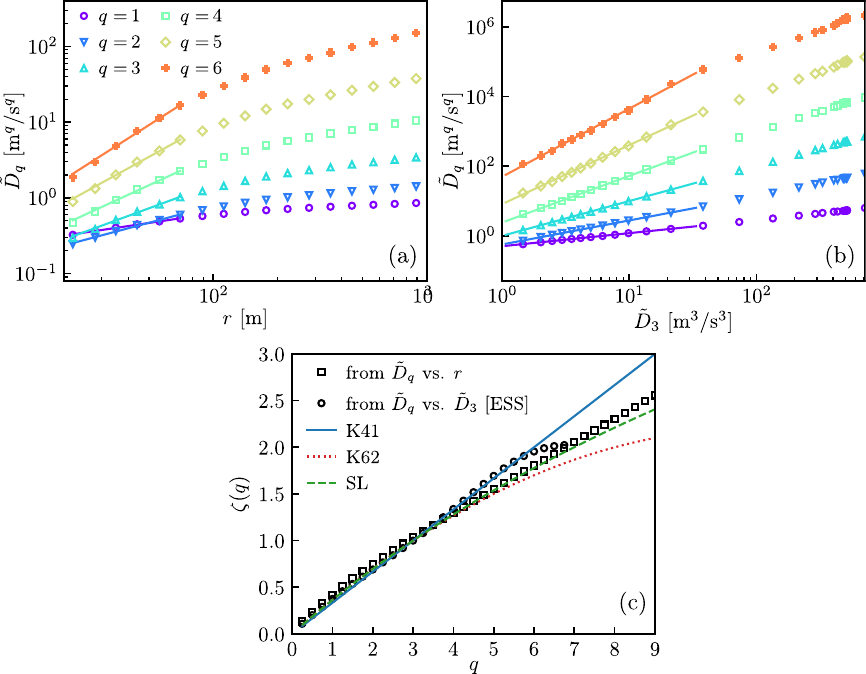}   
\end{center}
\caption{(a, b) Structure functions  $\tilde D_q$ (symbols) for absolute values of wind speed differences [see Eq.~\eqref{eq:structf_abs}] for height $h=90\,\si{m}$
as a function of (a) $r$ and (b) $\tilde D_3(r)$.
The solid lines are least-square fits to a power-law behavior in the inertial regime of 3D isotropic turbulence, see Eqs.~\eqref{eq:structf_scaling} and \eqref{eq:structf_ESSanalysis}.
(c) Exponents $\zeta(q)$ obtained from the power-law fits in (a) and (b). The dashed blue line referenced as K41 indicates a monofractal 
behavior with the exponent $\zeta(q)=q/3$, which applies when assuming a self-similar structure of the turbulent wind field \citep{Kolmogorov:1941a}. 
The red solid line indicates the nonlinear multifractal behavior of $\zeta(q)$ according to
Eqs.~\eqref{eq:zetaq_Kolmogorov} with $\mu=0.45$, referenced as K62. The green solid line refers to the nonlinear multifractal behavior
according to Eq.~\eqref{eq:zetaq_SheLeveque}, referenced as SL.} 
\label{fig:zetaq} 
\end{figure}
%--------------------------------------------------------------------------------------------------

\subsection{Scaling behavior of moments of horizontal wind speed increment distributions}     
For Taylor distances $r\lesssim h$, i.e.\ in the inertial regime of 3D isotropic turbulence, intermittency shows up in
large values of wind speed increments that are much more frequent than expected from Gaussian statistics.
The intermittency is reflected in multifractal behavior, given by the nonlinear $q$-dependence of exponents $\zeta(q)$ in Eq.~\eqref{eq:structf_scaling}.
Figure~\ref{fig:zetaq} shows $\tilde D_q$ as function of (a) $r$ and (b) $D_3(r)$ [ESS analysis] for the measurement height 
$h=90$\,m and various orders $q$. 
The solid lines in the figures
are least-square fits of power laws
to the data in the inertial regime. In agreement with results reported earlier in the literature \citep{Benzi/etal:1993}, 
the range of the power-law behavior  in the ESS analysis extends, corresponding to distances $r\gtrsim h$
[Fig.~\ref{fig:zetaq}(b)].

The slopes of the lines in the double-logarithmic plots yield
the exponents $\zeta(q)$, see Eqs.~\eqref{eq:structf_scaling} and \eqref{eq:structf_ESSanalysis}.
They are shown in Fig.~\ref{fig:zetaq}(c). At small $q$, the exponents approximately follow 
the linear monofractal behavior $\zeta(q)=q/3$ according to K41 scaling, which is indicated by the straight solid blue line. 
At larger $q$, the concave curvature of the $\zeta(q)$ signatures nonlinear multifractal behavior due to intermittency corrections.
The theoretical prediction \eqref{eq:zetaq_SheLeveque} [SL] describes quite closely
the exponents $\zeta(q)$
obtained from power-law scaling of $\tilde D_q$ with $r$.
However, neither of the Eqs.~\eqref{eq:zetaq_Kolmogorov} [K61] 
and \eqref{eq:zetaq_SheLeveque} [SL] gives $\zeta(q)$ accurately.

\section{Conclusions}
\label{sec:conclusions}
We have analyzed offshore horizontal wind speeds sampled in the North Sea at heights $30-90\,\si{m}$ above the sea level over a time period of 20~months with a time resolution of one second. 
Distributions and correlation properties were investigated for wind speeds and wind speed increments.
Cross-correlation functions were determined with respect to time and height differences. For equal-time wind speed differences at different heights, we carried out a detrended fluctuation analysis. 
In the analysis of distributions of wind speed increments and their moments, we considered a mapping of time lags $\tau$
onto spatial distances $r$ by applying the Taylor's hypothesis locally,
i.e.\ by using the mean wind speed $\bar u_\tau$ in every time window given by the lag $\tau$. We call the resulting $r=\bar u_\tau \tau$ Taylor distance. 

Distributions of horizontal wind speeds can be described by the Weibull form. This form, however, underestimates the frequency of high wind speeds
$\gtrsim 18\,\si{m/s}$ in the tail region.
Both the shape parameter $k$ and the scale parameter $\lambda$ 
of the Weibull distribution vary weakly with the height $h$ above the sea level. The shape parameter $k$ is about two and decreases logarithmically with $h$. The scale 
parameter $\lambda$ decreases approximately linearly with $k$. These results yield a height-independent turbulence intensity of about one half and a logarithmic increase of the mean wind speed with $h$, in agreement with previously 
reported findings and theories of wall turbulence. As a measure of the differences between wind speed distributions at different heights, we calculated
the symmetrized Kullback-Leibler divergence. It increases as $\sim\ln^2(h/h_0)$ with $h$, where $h_0$ is a reference height. 

Cross-correlation functions between wind speeds at different heights decay very slowly. Relative deviations between them
are significant for small times, while for large times $t\gtrsim 10^4\,\si{s}$ they are almost equal.
A detrended fluctuation analysis of differences between equal-time wind speeds at different heights showed a short- and long-time scaling regime
with Hurst exponents of about 0.75 and close to one. These large Hurst exponents 
reflect strong persistent correlations in the temporal changes of equal-time wind speed differences.

Cross-correlation functions between time derivatives $\partial_t u(t,h)$ of wind speeds, numerically represented by wind speed increments in the smallest time
interval $\Delta t=1\,\si{s}$, show anticorrelations. They slowly decay towards zero from negative values.
At short times, cross-correlations between $\partial_t u(t,h)$ and $\partial_t u(t,h + \Delta h)$ quickly drop to negative values and their magnitudes become much smaller 
for larger height differences $\Delta h$. For larger times $t\gtrsim 10^2\,\si{s}$, no significant variation with the vertical separation $\Delta h$ occurs.
Both the auto- and the even parts of cross-correlation functions satisfy sum rules required by the fact that the mean kinetic wind energy should not diverge, i.e.\ by the finite variance 
of wind speed distributions. Equal-time correlations between wind speed increments of time lag $\tau$ at different heights 
decay exponentially with the height separation $\Delta h$, where
the correlation length $\xi$ of the exponential decay increases monotonically with $\tau$.

Moments of the distribution of wind speed increments show intermittency corrections in the regime of 3D isotropic turbulence.
The intermittency is reflected in a multifractal scaling behavior of the moments with the Taylor distance. Models suggested for the multifractal scaling 
give an approximate but not perfect match to the data. 

Distributions of wind speed increments $\Delta_r u$ at 
Taylor distances $r$ have a tent-like shape in the core part around $\Delta_r u$ for small $r$.
With increasing $r$, their shape in the core part changes and approaches a Gaussian for large $r\gtrsim10^3\,\si{km}$. 

A striking feature in the distribution of  wind speed increments is the occurrence of a peak in their left tails 
for Taylor distances in the range $10-200\,\si{km}$. In this regime, turbulent wind fields are commonly believed to be governed by gravity waves.
Interestingly, the expected and observed linear scaling $D_3(r)\sim -r$ of the third-order structure breaks down in this regime, 
when the peak in the left tails is ignored, i.e.\ when using the shape of the core part to represent the full distributions including their tails.
This suggests that the peak is an intrinsic feature of atmospheric turbulence. It will be interesting to see whether further investigations 
can corroborate or confirm this finding. 

An open question concerns the dependence of our findings on varying stratification conditions.
In a recent combined analysis of
airborne and stationary measurements at the FINO1 mast \citep{Platis/etal:2022}, atmospheric stability was found
to depend on the measurement height $h$, where predominantly stable conditions were reported for $55\,\si{m}<h<95\,\si{m}$.
For quantifying atmospheric stability,
suitable criteria need to be considered based on, for example, the
lapse rate, the bulk Richardson number, and the Monin-Obukhov parameter \citep{Stull:1988}. Evaluation of these parameters
requires to include additional information about temperature and pressure profiles, air humidity and vertical wind velocities
in the analysis, which is beyond the scope of this study.

Our present analysis of offshore wind speed data allows for testing theoretical approaches against empirical findings and
thus lays a suitable basis to benchmark data obtained from stochastic wind speed modeling.
\vspace*{-2ex}

\begin{acknowledgements}
We thank J.~Peinke for valuable discussions and M.~W\"achter for helping us with the data acquisition. 
We are grateful to the BMWI (Bundesministerium für Wirtschaft und Energie) and the PTJ
(Projekttr\"ager J\"ulich) for providing the data of the offshore
measurements at the FINO1 platform. We acknowledge use of a high-performance computing cluster funded by 
the Deutsche Forschungsgemeinschaft (Project No.\ 456666331).
\end{acknowledgements}

%\bibliographystyle{spbasic_updated}
%\bibliography{Wind_analysis_short.bib}

%Control: production of eprint (1) enabled
%

\end{document}